\documentstyle[preprint,eqsecnum,aps]{revtex}
\begin{document}
\title{A Plane of Weakly Coupled Heisenberg Chains:
Theoretical Arguments and Numerical Calculations}
\author{Ian Affleck}
\address{Canadian Institute for Advanced Research and Physics Department,
University of British Columbia,\\ Vancouver, British Columbia, Canada V6T 1Z1}
\author{Martin P. Gelfand}
\address{Department of Physics, Colorado State University,
Fort Collins, Colorado 80523}
\author{Rajiv R. P. Singh}
\address{Department of Physics, University of California,
Davis, California 95616}
\maketitle
\begin{abstract}
The $S=1/2$, nearest-neighbor, quantum Heisenberg antiferromagnet 
on the square lattice with spatially anisotropic couplings is reconsidered, 
with particular attention to the following question: at $T=0$, does N\'eel order
develop at infinitesimal interchain coupling, or is there a nonzero
critical coupling?  A heuristic renormalization group argument is presented
which suggests that previous theoretical answers to that question are
incorrect or at least incomplete, and that the answer is not universal but
rather depends on the microscopic details of the model under consideration.
Numerical investigations of the nearest-neighbor model are carried out
{\it via\/} zero-temperature series expansions about Ising and dimer
Hamiltonians.  The results are entirely consistent with a vanishing
critical interchain coupling ratio $R_c$; if $R_c$ is finite, it is unlikely to
substantially exceed 0.02.

\hbox{}

\hbox{}


\hbox{}

\hbox{}

\noindent Short title: A Plane of Weakly Coupled Heisenberg Chains
\end{abstract}
\vfil\eject



\section{Introduction}
Lately there has been considerable interest in one-dimensional to
two-dimensional crossover in Luttinger liquids. The motivation is
largely the interest in possible non-Fermi liquid behavior in two
dimensional models for high-$T_c$ superconductors \cite{Anderson}.
While much of this work has focused on the Hubbard or $t$-$J$ model, a
simpler case to understand is that of the spatially anisotropic 
square-lattice spin-1/2 Heisenberg antiferromagnet,
\begin{equation} 
\label{eq:Ham}
H = \sum_{{\bf r}_i-{\bf r}_j = e_x}
J_x\,{\bf S}_i\cdot {\bf S}_j
\quad + \sum_{{\bf r}_i-{\bf r}_j = e_y}
J_y\,{\bf S}_i\cdot{\bf S}_j,
\end{equation}
(and see Fig.~\ref{fig:lattice}) with $0<J_y \ll J_x$. 
The essential question is whether N\'eel order sets in for
infinitesimal 
\begin{equation}
R \equiv J_y/J_x 
\end{equation}
or whether there is a finite minimum
coupling ratio necessary for long range order (LRO). 
A subsidiary question is the nature of the magnetically disordered
phase at small $R$, if the latter scenario holds true.
While various analytical and numerical investigations of this question have
appeared in the literature \cite{Sakai1,Azzouz1,Parola1},
it appears to be far from settled.

The purpose of this paper is twofold.  In Sec.~\ref{sec:th_args}
we review previous analytic arguments for or against the existence of a
finite minimum ratio.  We then discuss a renormalization group
framework within which previous arguments appear to be
incorrect or incomplete. This approach indicates
that the behavior is not universal and that for any particular
model the question may only
be answered by numerical investigation.  Therefore in Sec.~\ref{sec:ser_calc}
we present such a numerical investigation of the Hamiltonian (\ref{eq:Ham})
based on high order series expansion about both the N\'eel and dimer phases.
That analysis seems to indicate that if there is a critical ratio it is
rather small.

\section{Theoretical Arguments}
\label{sec:th_args}
\subsection{Spin-Wave Theory}
\label{sec:swt}
Spin-wave theory plays an important role in our understanding of
antiferromagnetism.  A systematic $1/S$ expansion (where $S$ is the
spin magnitude) is generally believed to give, in low orders, a
fairly accurate estimate of the sublattice magnetization on various
lattices.  In particular, it correctly predicts the absence of long
range order at $T=0$ for a one-dimensional antiferromagnet.  This
effect can be seen from the leading order correction to the
sublattice magnetization.  For the anisotropic 2D model of Eq.~(\ref{eq:Ham}),
we have: \begin{equation}
\langle S^z_i \rangle = \pm \left[ S-{1\over2}\left( 
\int{d^2{\bf k}\over (2\pi)^2}{1\over \sqrt{1-\gamma_{\bf k}^2}}-1 \right)
\right],\label{eq:spinwave}\end{equation} where
\begin{equation} \gamma_{\bf k}\equiv {J_x\cos k_x+J_y\cos k_y\over
J_x+J_y}\end{equation} and the $\bf k$ integral runs over the Brillouin
zone, $|k_i|<\pi$. For $J_y=0$, the integrand is independent of $k_y$
and the integral has logarithmic divergences at $k_x=0$ and $\pi$
indicating the absence of N\'eel order.  For finite $J_y$ the integral
is finite and for small $R$ we obtain:
\begin{equation} \langle S^z_i \rangle \approx \pm \left[ S+{1\over \pi}
\ln R \right].\end{equation}  
If we took as a criterion for the stability of
the N\'eel phase that the $O(1)$ correction should be smaller than the
leading $O(S)$ term, then we would conclude that N\'eel order breaks
down at:
\begin{equation} R_c\approx e^{-\pi S}.\end{equation}  For $S=1/2$,
numerical evaluation of the integral in Eq.~(\ref{eq:spinwave})
gives $R_c\approx .03367$ \cite{Sakai1}.

\subsection{Chain Mean Field Theory} 
\label{sec:cmft}
A standard method for treating dimensional crossover problems
of this type is to use the known behavior of the antiferromagnetic
susceptibility for the one-dimensional system and to treat the
couplings in the second dimension in mean field theory \cite{Scalapino1}.
(We refer to this as ``chain'' mean field theory to distinguish it from
another mean field theory to be discussed later.)
The Hamiltonian of Eq.~(\ref{eq:Ham}) is replaced by the following mean field
Hamiltonian for each chain: 
\begin{equation}
H_{CMF} \equiv \sum_i J_x {\bf S}_i\cdot {\bf S}_{i+1} +
2J_y \langle S^z\rangle \sum_i (-1)^i S^z_i.\end{equation}  
We now calculate $\langle S^z\rangle$ using
this Hamiltonian and demand self-consistency.  The mean field critical
point is then determined by:
\begin{equation} 1=2R\chi_1(T_N),\label{J_cMF}\end{equation} where
$\chi_1$ is the zero-frequency antiferromagnetic
susceptibility for the one-dimensional chain with $J_x=1$.  
For a half-integer spin Heisenberg antiferromagnet,
$\chi_1(T)$ diverges as $1/T$, as $T\to 0$.  This argument then
predicts a finite N\'eel temperature: \begin{equation} T_N \approx
J_y.\end{equation}  No matter how small $J_y$, the ground state is always
ordered.  On the other hand, for integer $S$, $\chi_1(0)$ has a finite
value of order $1/\Delta$, the Haldane gap.  Hence this argument
predicts a disordered ground state for integer $S$ and $R<R_c$ with
$R_c\approx \Delta$. This result is readily generalized to
$d$-dimensional systems.  The factor of $2$ in Eq.~(\ref{J_cMF}) is simply
replaced by the number of nearest neighbor chains.  Indeed, it
presumably becomes exact in the limit $d \to \infty$.  It has an
obvious problem for $d=2$ where the Mermin-Wagner theorem tells us that
N\'eel order should be impossible at any finite temperature. 
Nonetheless, we might be tempted to believe the conclusion that it
does occur at $T=0$.  A more sensible application of this mean field
theory for $d=2$ is to the case of a staggered intra-chain coupling 
at $T=0$ (see Sec.~\ref{sec:dimerexp}, below).  
Eq.~(\ref{J_cMF}) then becomes:
\begin{equation} 1=2R\lambda^c_{\rm CMF}\chi_1(\lambda^c_{\rm CMF}).
\end{equation}  
Here $\lambda$ is the ratio of coupling on alternate links, and
$\lambda^c_{\rm CMF}$ is its critical value within the chain mean field theory.
Since a staggered interaction in one dimension has scaling dimension $1/2$, we
conclude that $\chi_1(\lambda ) \propto (1-\lambda )^{-2/3}$ (up to log
corrections).  Now $1-\lambda$ plays a role roughly analogous to a
finite temperature; as $\lambda \to 1$, the mean field theory predicts
that N\'eel order sets in at $1-\lambda^c_{\rm CMF}\propto R^{3/2}$.

\subsection{Renormalization Group Argument}
Consider a system of  quantum  chains weakly coupled to each
other at $T=0$.  They may be of $XY$ or Heisenberg symmetry. 
Alternatively, consider a system of classical spins consisting of
planes which are weakly coupled to each other, at finite $T$.  In
either case (ignoring topological terms, for the moment), we may
represent the system at long length scales by a 3-dimensional
non-linear $\sigma$-model, with action or Hamiltonian: \begin{equation}
S=(\Lambda /2g )\int d^3 {\bf x} [(\partial_x\vec \phi
)^2+(\partial_y\vec \phi )^2+ R (\partial_z\vec \phi
)^2]\end{equation} Here $\vec \phi$ has unit length and either 2 or 3
components in the $XY$ or Heisenberg case.  As before, $R$
is the ratio of inter-chain (or plane) coupling to in-chain
(or plane) coupling.  $\Lambda$ is the ultraviolet cutoff; {\it i.e.}, the
field, $\vec \phi$ has Fourier modes with $|k_i|<\Lambda$.  (We cut off
the momentum inside a cube.)  The factor of $\Lambda$ is  inserted to
make the three-dimensional coupling constant $g$ dimensionless, as is
usually done in formulating the 3D renormalization group (RG) equations.  

Note that the only trace of the quasi-2-dimensionality is the
anisotropy in the $( \partial \vec \phi )^2$ terms.  We can get rid of
this by a rescaling of $z$ by a factor of $\sqrt{R}$, that is, we
define a new length co-ordinate, $z^{'}\equiv z/\sqrt{R}$.  In
momentum space, we define a new 3-component of momentum, $k_3^{'}\equiv
\sqrt{R}k_3$.  The action now looks completely 3-dimensional
except that the cut-off is no longer a 3D cube but a squat box
of area $\Lambda^2$ and height $\sqrt{R}\Lambda$.  To complete
the elimination of the anisotropy from the action, we reduce the length
and width of the box to $\sqrt{R}\Lambda$ also, by integrating
out higher momentum modes, the standard RG procedure.  Because the
inter-chain coupling is so weak, this may essentially be done using the
2D RG equations;  in momentum space, $(k^{'}_z)^2$ is so small (due
to the small cut-off) that we consider the planes to be essentially
decoupled until we have lowered the 2D cutoff down to approximately
$\sqrt{R}\Lambda$.  The value of the 2D effective coupling, when
the cut-off has been reduced to $\sqrt{R}\Lambda$, then acts as
the initial condition for further RG calculations:  as we lower the
cut-off still further we should use the isotropic 3D RG equations using
$g_2(\sqrt{R}\Lambda )$ as the initial value.  In the limit
$R \to 0$, the 2D coupling flows to its zero-cutoff fixed
point, $g_2(0)$.  
See Fig.~\ref{fig:rgflow} for a sketch of one possible flow diagram.

In both $XY$ and Heisenberg cases, the 3D system has ordered and
disordered phases, separated by some critical coupling $g_c$. (For
$g<g_c$ the system is ordered.)  Thus, whether or not the system orders
for arbitrarily small $R$ is determined by whether or not
$g_2(0)<g_c$.  In the case of Heisenberg symmetry with no topological
term, $g_2(0)$ is naively infinite, so the 3D system is in the
disordered phase for sufficiently small $R$.  
From another point of view, the 2D system develops a
finite correlation length, $\xi$, as we reduce the cut-off.  For
sufficiently small $R$, $ \sqrt{R}\Lambda  < 1/\xi$. 
Then  further renormalization using the 3D RG equations cannot
eliminate this finite correlation length.  In the $XY$ case $g_2(0)$ is
basically the renormalized dimensionless temperature.  It has a finite
value along the Kosterlitz-Thouless (KT) critical line, $g_2(0)<g_{KT}$. 
The important question is whether or not $g_{KT}<g_c$.  If it is, then
an arbitrarily weak inter-plane coupling leads to order for all
$T<T_{KT}$.  In the other case, if $g_c<g_{KT}$, then there will be
some special temperature, $T_c<T_{KT}$, such that below this
temperature an arbitrarily weak inter-plane coupling leads to LRO but
above this temperature there is a minimum inter-plane coupling
necessary for LRO.  

Numerical simulations of the classical layered $XY$ model
indicate that $g_{KT}<g_c$ \cite{Chui}.  In
this case the present approach predicts a critical coupling,
$g_c(R)$, for  weakly coupled planes, slightly above
$g_{KT}$.  This can be estimated as the bare coupling for which the
renormalized coupling at scale $\sqrt{R} \Lambda$,
$g_2(\sqrt{R} \Lambda)\approx g_c$.  Using the 2D Kosterlitz
renormalization group equations this gives: 
\begin{equation}
g_c(R)-g_{KT} \propto 1/(\ln R )^2,\end{equation} 
in agreement with estimates based on plane mean field theory \cite{Chui}
and another approach \cite{Hikami,Kosterlitz} somewhat closer in
spirit to the present one.

Now let us consider the half-integer spin quantum Heisenberg
chain \cite{Affleck1}.
The two-dimensional $\sigma$-model action  now has an extra
topological term added to it, with topological angle $\pi$.  This
angle itself does not renormalize, for symmetry reasons, but it has
a crucial effect on the renormalization of $g_2$.  There is now a
finite coupling critical point, $g_2(0)$,  shown in Fig.~\ref{fig:rgflow}.
(The strong coupling phase is spontaneously dimerized; it can be
reached with a sufficiently strong antiferromagnetic second nearest
neighbor coupling.) The three-dimensional RG
flows are expected to be the same for both integer and half-integer
spin, although the nature of the disordered phase may depend on
whether the spin is half-integer, odd integer or even integer, due
to a Berry's phase topological term \cite{Haldane}.  We assume the
only important effect of the topological terms is to produce a
finite $g_2(0)$ critical point in two dimensions, as 
shown in Fig.~\ref{fig:rgflow}.  The situation is then similar  to the
classical $XY$ case.  Whether or not a disordered phase occurs for
weak interchain coupling depends on whether or not $g_2(0)>g_c$.

Let us contrast this approach to either of those 
discussed above, in Secs.~\ref{sec:swt} and \ref{sec:cmft}.
First consider the
calculation of the breakdown of the N\'eel state in lowest order
spin-wave theory.  This can be regarded as the above calculation of the
renormalization of the 2D coupling, using {\it only the lowest order 2D
RG equations}.  For large $S$ we begin with a very small bare coupling.
However, the lowest order RG equations always break down for small
enough $R$.  Furthermore, these equations do not distinguish
integer and half-integer $S$.  In $\sigma$-model language, the lowest
order RG is independent of the topological term.  Actually the RG
equations are independent of it to all orders in $g$; its effects are
exponentially small in $g$.  Nonetheless, for small enough $R$
we always renormalize into the regime where these non-perturbative
contributions are important.  Blind use of the lowest order equations
essentially leads to the conclusion that $g_2(0)$ is infinite and hence
that the system is disordered at small enough $R$.

On the other hand consider the standard mean field argument \cite{Scalapino1}.
Because the 2D susceptibility is
divergent along the whole KT critical line, or in the Heisenberg case
with topological term, it predicts LRO in all these cases.  We
argue that in order to go beyond a mean field treatment one should
consider not the 2D susceptibility but rather the renormalized 2D
effective coupling.  Naively if the 2D susceptibility is infinite, the
2D effective coupling is zero.  But in fact, this is not actually the
case.  Since the 2D effective coupling is finite at zero cut-off we
must consider whether or not it is below $g_c$, leading to the above
conclusion.

It was suggested recently by Parola {\it et al.\ }\cite{Parola1}
that, for $R$ less than a finite $R_c$, the long wavelength behavior
``can be interpreted in terms of
decoupled one dimensional chains.''  The implausibility of this
proposal can be seen by considering a finite number of chains.  In
this case, we can analyze the scaling behavior entirely in terms of
the (1+1)-dimensional   renormalization group.  The inter-chain
coupling, of dimension 1, is relevant.  We might then naively expect
from standard scaling arguments that for two chains there should be a gap,
proportional to $R$ (up to logarithms).  The existence of a gap in this
case has been shown numerically \cite{twochain_num,White1}.
In the case of three chains, White {\it et al.\ }\cite{White1} found that the 
gap appears to vanish and
this has been argued to be the case for any odd number of chains. 
However, this does not necessarily mean that the system asymptotically
behaves like decoupled chains.  Indeed, it seems more likely that the
low energy states would be those of a {\it single\/} $S=1/2$ chain. 
This assertion is motivated by the behavior of a single spin $S$ chain 
as a function of $S$.  
For integer $S$ one expects a gap while for half-integer $S$ one
expects universal gapless behavior which is independent of $S$ ({\it i.e.},
always in the free boson or $k=1$ Wess-Zumino-Witten universality
class).  For half-integer $S$ the density of low-lying excitations
(after scaling out the spin-wave velocity) {\it does not increase\/}
with increasing $S$.  We may expect similar behavior for $2S$ coupled
spin-$1/2$ chains:  inter-chain coupling is relevant and the system
scales away from decoupled chains to the single chain fixed point
with fewer gapless degrees of freedom.  
Note that in this case, the
number of low-lying degrees of freedom does not scale with the area of
the system but only with the length.
It is of course
possible that, as the number of chains is increased, the gap to other
excitations decreases and asymptotically approaches zero as the number
of chains goes to $\infty$.  Indeed, this must happen if the
$\infty$-chain system N\'eel orders.  However, it then seems unlikely
that the low-lying excitations would be those of decoupled chains.  We
rather expect that, if a disordered phase exists at small $R$, it has
a genuinely 2-dimensional nature.  Previous work on the isotropic
$S=1/2$ square lattice Heisenberg antiferromagnet
with next-nearest neighbor interactions suggests
that a dimer-ordered ground state is the most likely to occur in a
magnetically disordered phase \cite{frustratedAF}.
However, other possibilities with or without
a gap cannot be ruled out.

Actually, the result of our reasoning is pretty
uninformative.  It is simply that we cannot tell from RG arguments
whether or not the system orders for arbitrarily weak inter-plane
coupling, in cases where the 2D system has an infinite correlation
length. We might expect that the existence or non-existence of a
critical $R$ is not universal.  Different realizations of the model
may have different values of $g_c$ or $g_2(0)$, since critical
temperatures (or couplings) are in general not universal.  For
instance, a different phase diagram might ensue if the chains are
coupled with a first and second nearest neighbor inter-chain coupling,
both proportional to $R$. It would thus seem that the question of the
existence of a disordered phase for finite $R$ in a particular model
must be answered by numerical work.  

\section{Series Calculations and Analyses}
\label{sec:ser_calc}
We now turn to numerical studies of weakly coupled, $S=1/2$
Heisenberg chains, and in particular the anisotropic two-dimensional
lattice of Fig.~\ref{fig:lattice} and Eq.~(\ref{eq:Ham}).
It is obviously not possible for any numerical calculation to
distinguish between $J_y^c=0$, {\it i.e.,} the persistence of
N\'eel order for arbitrarily small interchain couplings, and
a sufficiently small but positive $J_y^c$.
Provided that there is no strong evidence for any particular
nonzero value of $J_y^c$, the best one can do is argue that
numerical data are consistent with a vanishing $J_y^c$ and
offer reasonable upper bounds on its value.

Our numerical studies consist of a variety of zero-temperature
series expansions ({\it i.e.,} Rayleigh-Schr\"odinger perturbation theory)
following the cluster-expansion techniques described in 
Ref.~14. 
Series expansions have several advantages over finite-size
calculations in two-dimensional lattice quantum many-body problems.
There is no need to worry about cluster-shape effects --- which
should be of particular concern in anisotropic models such as
the one of present interest.
Given comparable computing power, series calculations can account for much 
further-range correlations than would be possible in an exact diagonalization
calculation (see the discussion in Sec.~V of Ref.~14). 
Finally, the fact that the series calculations do not directly study
the model of interest, but rather yields results on a one (or more)
parameter family of models, may allow for further informative comparisons
with approximate analytic calculations.

Expansions were carried out about both Ising and dimer Hamiltonians.
The calculations and analyses will be described in detail below; here
let us preview the results.
The results of both types of expansions are consistent with $J_y^c=0$.
We believe that $J_y^c$ is unlikely to exceed $0.02\,J_x$.
This upper bound is not much smaller than the lowest-order
spin-wave estimate (and hence by itself cannot be taken as very strong
evidence that spin-wave theory is qualitatively wrong);
however it is notably less than the value $0.1\,J_x$
suggested by Parola {\it et al.\ }\cite{Parola1} based on
exact diagonalization of clusters with up to 32 spins.
In addition, the Ising expansions yield estimates for the
staggered magnetization and the correlation-length anisotropy
which are in excellent agreement with spin-wave theory
for $J_y/J_x$ down to 0.1.
Finally, the dimer expansions appear to be consistent with
the chain mean field theory of Sec.~\ref{sec:cmft} at small
$J_y$, and thus support the proposition that $J_y^c=0$.
However, that conclusion must be tempered by comparison of the dimer
expansions for a plane of chains with the corresponding calculations
for a {\it pair\/} of chains.

\subsection{Ising expansions}

In order to discuss the Ising expansions, we consider a generalization
of the coupled-chain Hamiltonian (\ref{eq:Ham}) in which Ising
anisotropy is introduced, namely 
\begin{eqnarray}
\nonumber H &=& \sum_{{\bf r}_i-{\bf r}_j = e_x}
J_x[S_i^z S_j^z +\alpha(S_i^xS_j^x+S_i^yS_j^y)] \\
& &+ \sum_{{\bf r}_i-{\bf r}_j = e_y}
J_y[S_i^z S_j^z +\alpha(S_i^xS_j^x+S_i^yS_j^y)].
\end{eqnarray}
Henceforth we choose units of energy so that $J_x=1$ (and $J_y=R$).
Physical quantities are expanded in powers of $\alpha$;
we have calculated the ground state energy
$E_g$, the sublattice magnetization $M=\langle S_0^z \rangle$,
and the correlation length anisotropy $(\xi_y/\xi_x)^2$ which is
given explicitly by
\begin{equation}
\sum_i \langle S_0^z S_i^z \rangle_c y_i^2 \Big/ \sum_i \langle S_0^z S_i^z\rangle_c x_i^2,
\end{equation}
where the subscript $c$ refers to connected correlations
\begin{equation}
\langle S_0^z S_i^z \rangle_c =
\langle S_0^z S_i^z \rangle-\langle S_0^z \rangle \langle S_i^z \rangle.
\end{equation}

The energy and magnetization series were determined to order
$(\alpha^2)^5$, while the $(\xi_y/\xi_x)^2$ series were only
calculated to order $(\alpha^2)^4$.
Note that each value of $J_y$ requires a separate calculation
of the series; since it is not possible to present the complete
series expansions in a compact format they are not displayed
here, but are available as supplementary material.
Here we discuss the analysis.

The energy series were analyzed by direct Pad\'e approximants.
The results of five different Pad\'e approximants evaluated at $\alpha=1$
are shown in Fig.~\ref{fig:I_E_pade}.
It is clear that they extrapolate smoothly between the
one and two dimensional limits, that is to say between $J_y=0$
and $J_y=1$.
This is consistent with the expansion being convergent up to
$\alpha=1$ for all $J_y \geq 0$, which is in turn consistent 
with $J_y^c=0$.
However, this is quite weak evidence, as we will discuss in
connection with the dimer expansions.
Perhaps the results for $E_g$ could serve best as a touchstone
for the quality of finite-size calculations.

To analyze the magnetization series, we first make a change of variables
originally introduced by Huse, $\delta=1-(1-\alpha^2)^{1/2}$,
which removes the square-root singularity at $\alpha=1$ expected on the
grounds of spin-wave theory.
The results for five Pad\'e approximants evaluated at $\alpha=1$
are shown in Fig.~\ref{fig:I_M_pade}.
For $J_y \geq 0.2$ the approximants are well-converged and appear
to be in good agreement with spin-wave theory.
For $J_y < 0.2$ the convergence is poor, but the approximants are
suggestive of N\'eel order for all $J_y$  and with $M$ vanishing as a
small power of $J_y$.
Note that the chain mean field theory implies
$M \sim J_y^{1/2}$ as $J_y\to0$.
In any case, these data provide no evidence that $M\to0$ at any
particular $J_y^c>0$.

The correlation length anisotropy series are short, however this
quantity is nonsingular as $\alpha\to1$ and the Pad\'e approximants
evaluated at $\alpha=1$ are extremely consistent 
even down to very small $J_y$: see Fig.~\ref{fig:I_aniso_pade}.
For $J_y \geq 0.1$ there is 
remarkable agreement between the series estimates and the
spin-wave calculation presented by Parola {\it et al.\ }\cite{Parola1}
For smaller $J_y$, two of the three approximants indicate
$\xi_y/\xi_x \to 0$ for $J_y$ between 0 and 0.02, and the other
approximant is ill-behaved.
These data are consistent with $J_y^c=0$, but also with a 
small critical interchain coupling.

\subsection{Dimer expansions}
\label{sec:dimerexp}
The dimer expansions are in the variable $\lambda$
for properties of the Hamiltonian $H=H_0 + \lambda H_1$,
where for $H_0$ we take the columnar dimer Hamiltonian
\begin{equation}
H_0 = \sum_{\langle ij \rangle \in \cal D} J_x\ {\bf S}_i\cdot {\bf S}_j,
\end{equation}
with $\cal D$ the dimer covering of the square lattice shown 
in Fig.~\ref{fig:dimercov}, and $H_1$ is the remainder of the couplings, 
so that $\lambda=1$ corresponds to the coupled-chain model of interest, namely
\begin{equation}
H_1 = \sum_{{\bf r}_i-{\bf r}_j = e_x; \langle ij \rangle \not\in \cal D}
J_x\ {\bf S}_i\cdot {\bf S}_j
+ \sum_{{\bf r}_i-{\bf r}_j = e_y} 
J_y\ {\bf S}_i\cdot {\bf S}_j.
\label{eq:dimerexpHam}
\end{equation}

The quantities for which we have obtained series expansions, 
to order $\lambda^7$ (which involves evaluation of 1041 graphs),
include the ground state energy $E_g$, moments of the antiferromagnetic 
equal-time structure factor
\begin{equation}
M_1 = \sum_i \langle {\bf S}_0\cdot {\bf S}_i \rangle \pi({\bf r}_i),
\end{equation}
\begin{equation}
M_{rr}= \sum_i \langle {\bf S}_0\cdot {\bf S}_i \rangle \pi({\bf r}_i)r_i^2,
\end{equation}
(where $\pi({\bf r}_i)$=1 if ${\bf r}_i$ lies on 
the same sublattice as ${\bf r}_0$, and is $-1$ otherwise),
and the antiferromagnetic susceptibility $\chi$.

It should be clear from the form of the Hamiltonian that all these
quantities possess two-variable expansions, in powers of $\lambda$ and
$\lambda J_y$.  
Hence the coefficient of $\lambda^m$ in any of these series can be
expresses in terms of an order-$m$ polynomial in $J_y$;
and calculations of the seventh-order series at eight values of
$J_y$ allows one to determine them at {\it any\/} $J_y$ by means
of polynomial interpolation. 

Before considering the details of the dimer series analysis, let us look at the
notions underlying this approach, and what we might expect to learn from it.
Properties of the N\'eel-ordered phase are inaccessible to the dimer
expansions (in contrast to the Ising expansions); what they can provide are
estimates of $\lambda^c(J_y)$, the smallest value of $\lambda$ for
a given $J_y$ at which the antiferromagnetic correlation length
and susceptibility diverge.
If one assumes the simplest possible phase diagrams
in the $J_y$-$\lambda$ plane, sketched in Fig.~\ref{fig:simplest_phase_diagrams},
then $\lambda^c(J_y)<1$ implies that the uniformly coupled system
($\lambda=1$) exhibits N\'eel order for that value of $J_y$.
Furthermore, one may compare the series estimates of $\lambda^c(J_y)$
with approximate analytic calculations for the critical line.

Let us first briefly discuss the ground state energy  series. 
Direct Pad\'e approximants allow for estimates of
$E_g(J_y,\lambda=1)$ which are shown in Fig.~\ref{fig:I_E_pade}.
The approximants are consistent with each other --- and with the
values obtained from the Ising expansions for $J_y$ as large as 0.4.
This implies that the singularity in $E_g$ at the critical line
is very weak; the alternative, that $\lambda^c>1$
for $J_y<0.4$, is ruled out by both the Ising expansions and
the other dimer expansions.

The terms of the $M_1$, $M_{rr}$, and $\chi$ series are all positive
and increasing with order in $\lambda$.
Thus one may estimate $\lambda^c(J_y)$ either by ratio analyses
or inhomogeneous differential approximants (of which Dlog-Pad\'e
approximants are a special case) \cite{IDA}.

The estimates of $\lambda^c(J_y)$ which result from consideration
of $M_1$ are poorly converged, so we will only
discuss $M_{rr}$ and $\chi$.
The results for these two sets of series are presented in
Fig.~\ref{fig:D_planestuff}.
Let us address what the various curves and points signify.
First, six differential approximant estimates for $\lambda^c(J_y)$
are presented for several values of $J_y$; 
these particular approximants are chosen
because they utilize all the terms in the series and they yield
good estimates of $\lambda^c$ at both $J_y=0$ and $J_y=1$.
The missing approximants at any $J_y$ are either defective
or (in one or two cases) off-scale.
The approximants are reasonably consistent over the entire
range of $J_y$.
Second, the thin solid and long-dashed lines are the estimates of
$\lambda^c$ based on ratio analysis of the three highest-order
terms in the $\chi$ and $M_{rr}$ series.  
For a series $\sum_n c_n \lambda_n$, the ratios $c_n/c_{n-1}$
are plotted versus $1/n$ and pairwise linearly extrapolated to
$1/n=0$; the intercepts yield estimates of $1/\lambda^c$ based on
three consecutive terms of the series.
The $\chi$ ratio curve lies close to the differential
approximants, which reflects the fact that the $\chi$ series
are extremely well-behaved.
The ratio plot (corresponding to $J_y=1$) in Fig.~2 of Ref.~14 
provides further evidence that the antiferromagnetic susceptibility series
is better behaved than other dimer expansions.
Both the ratio and differential approximant analyses
suggest that $\lambda^c(J_y)<1$ for all $J_y>0$, and
hence that there is no magnetically disordered phase
along the line $\lambda=1$.

Third, two mean field estimates of $\lambda^c(J_y)$ are 
presented in the figure.  
The dotted-dashed line is based on ``dimer mean field theory,''
\cite{GSLDQAF} that is, the zeroth- and first-order terms
in $\chi(\lambda)$ are used to estimate $\lambda_c$.
(This is equivalent to considering a single dimer subject
to a staggered field which is determined, self-consistently,
by its staggered magnetization.)
That result is $\lambda^c_{\rm DMF}=1/(1+2J_y)$, and
it has the remarkable (and accidental) feature that it yields 
the exact $\lambda^c$ when $J_y=0$.
The broad solid line is based on the chain mean field theory
discussed in Sec.~\ref{sec:cmft}.
The values of $\chi_1(\lambda)$ (the antiferromagnetic susceptibility
for a single chain), which
are the essential input into that mean field theory,
were obtained by integrating a differential approximant to
the seventh-order series presented in Ref.~14. 
The broad solid line ends at $J_y\approx0.02$ because the numerical
estimates of $\chi(\lambda)$ are not reliable for $\lambda$ arbitrarily
close to 1.
Over most of the range of $J_y$ plotted, the dimer and chain mean field
theories give quite close values of $\lambda^c$.  The agreement is somewhat
better than one would expect: in both cases the leading behavior of $\lambda^c$
at large $J_y$ is $1/2J_y$, but the next-order terms are $-3/4J_y^2$ and
$-1/4J_y^2$ for the chain and dimer mean field theories, respectively,
and at $J_y=1$ the difference is hardly negligible!
At small $J_y$ the chain mean field theory is clearly in better agreement
with the series extrapolations than the dimer mean field theory.

To briefly recapitulate, the dimer series for the two-dimensional lattice of
coupled chains are entirely consistent with $J_y^c=0$.
However, there is reason to doubt the strength of this conclusion.
Since $\lambda^c(0)=1$ and $\lambda^c(1)\approx 0.54$, it would
be entirely natural to conclude from too-short series that 
$\lambda^c(J_y)$ interpolates smoothly between these two endpoints,
even if the correct result is that $\lambda^c$ does not exist for
$0 < J_y < J_y^c$ as would be the case if the scenario of 
Fig.~\ref{fig:simplest_phase_diagrams}(b) were to hold.
As a partial test of the reliability of the series estimates
of $\lambda^c(J_y)$, we have considered the problem of {\it two\/}
coupled chains by the dimer series expansion method.
The chains were taken to lie parallel to the bonds in the columnar dimer
configuration.  
The only differences between these calculations and those
preceding are the set of connected clusters (and their lattice constants),
and that we will present results for $M_{xx}$ rather than $M_{rr}$.
(In fact one can use $M_{xx}$ to estimate $\lambda^c(J_y)$ for the
plane of chains as well; the differences between the estimates based
on $M_{xx}$ and $M_{rr}$ are insignificant for the $J_y$ of interest.)

For two chains the only critical point in the $\lambda$-$J_y$ plane
is $(\lambda=1,J_y=0)$ \cite{twochains}.
Estimates of $\lambda^c(J_y)$ coming from inhomogeneous differential
approximants to the $\chi$ and $M_{xx}$ series for two chains
are displayed in Fig.~\ref{fig:twochains}, on the same scale 
as in Fig.~\ref{fig:D_planestuff} for ease of comparison.
The thick solid and thin dotted-dashed curves are the
chain and dimer mean field results, respectively, for
{\it two\/} chains.  (One obtains these from the
mean field results for planes of chains by the substitution
$2J_y \to J_y$.)
Several points are evident upon inspection.
First, even at very small $J_y$ one can distinguish between the
estimates of $\lambda^c$ from the differential approximants
for the plane of chains and for two chains; they seem
to approach zero with different slopes, on these plots.
This could be taken as further evidence that, for the plane of chains,
$J_y^c$ is extremely small if not vanishing.

However, for $J_y\alt 0.2$, where the approximants are well-converged,
one might conclude that there are critical points at $\lambda\approx 1$,
contrary to the known behavior of coupled pairs of chains.
We believe this line segment of ``pseudocritical'' points reflects
the existence of local maxima in the correlation length
at $\lambda\approx1$ for fixed, small $J_y$, where the
correlation length at these maxima exceeds the typical cluster
length in the seventh-order calculation (which is roughly 10).
That such pseudocritical points should exist is entirely plausible:
numerical studies for pairs of finite chains \cite{Parola1,twochain_num}
find that the gap initially {\it decreases\/} as $J_y$ is increased
from zero.
What is somewhat disturbing is that, for $J_y\alt0.1$, the two-chain 
approximants are as consistent with chain mean field theory (for two chains) 
as the plane-of-chain approximants are with chain mean field
theory (for a plane of chains).
Thus it is conceivable that for the plane of chains, the values of $\lambda^c$
indicated by the series at sufficiently small $J_y$ 
are pseudocritical points, rather than true critical points, as well.

To conclude, the problem of coupled chains at small
$J_y$ poses significant challenges to numerical studies.
Although the series expansions are consistent with $J_y^c=0$
they do not rule out a small but positive $J_y^c$.
If the disordered phase exists, the correlation length along the $x$ 
direction is probably large throughout that phase, and the 
gap is everywhere small.
A potentially fruitful avenue for future studies would be 
the consideration of models which include 
further-neighbor couplings; suitable models would have
larger critical interchain couplings within spin-wave theory
than the simplest coupled Heisenberg chain model studied here.

ACKNOWLEDGMENTS: IA thanks J. Fr\"ohlich, M. Gingras, and E. S\o rensen for 
helpful discussions. MPG thanks Keith Briggs, for sharing the code for his
differential approximant calculation program ``DA,'' and
D. Frenkel for discussions.
IA was supported in part by NSERC of Canada.
MPG acknowledges support from the MacArthur Chair at
the University of Illinois (and the hospitality of its holder,
A. J. Leggett) during the initial stages of this work.
RRPS was supported by NSF grant number DMR-9318537.

\begin{figure}
\caption{The spatially anisotropic Heisenberg model.
The solid and dashed segments correspond to interactions $J_x$
and $J_y$.}
\label{fig:lattice}
\end{figure}

\begin{figure}
\caption{
Renormalization group flows for the half-integer spin quantum 
Heisenberg models.
Here we assume $g_2(0)<g_c$ so the system orders for arbitrarily small $R$.
}
\label{fig:rgflow}
\end{figure}

\begin{figure}
\caption{Pad\'e approximants for the ground state energy, evaluated at $\alpha=1$.
The symbols are approximants to Ising series, while the lines
(which are connecting points with $J_y=0$, 0.1, 0.2, 0.3, 0.4, 0.5, and 0.75) 
are approximants to dimer series.}
\label {fig:I_E_pade}
\end{figure}

\begin{figure}
\caption{Pad\'e approximants to the Ising expansion for
the sublattice magnetization, evaluated at $\alpha=1$ following
the change of variables described in the text.
}
\label {fig:I_M_pade}
\end{figure}

\begin{figure}
\caption{Pad\'e approximants to the Ising expansion for
the correlation length anisotropy, evaluated at $\alpha=1$.
}
\label {fig:I_aniso_pade}
\end{figure}

\begin{figure}
\caption{The columnar dimer covering $\cal D$ used for 
the dimer expansions; see Eq.~(\protect\ref{eq:dimerexpHam}).}
\label {fig:dimercov}
\end{figure}

\begin{figure}
\caption{Sketches of the simplest plausible phase diagrams in the
first quadrant of the $J_y$--$\lambda$ 
plane for the Hamiltonian (\protect\ref{eq:dimerexpHam})
assuming that (a) $J_y^c=0$, and (b) $J_y^c>0$.
Note that units of energy are chosen so that $J_x=1$.
The phase diagrams must satisfy two contraints that follow
from the invariance of correlation functions with respect to 
multiplication of the Hamiltonian by a constant:
$(J_y,\lambda)=(a,1)$ and $(1/a,1)$ lie in the same phase,
and so must $(a,b)$ and $(ab,1/b)$.
In both cases the hatched regions constitute the N\'eel ordered
phase, which are surrounded by lines of critical points.
The one-dimensional Heisenberg critical point at $(J_y,\lambda)=(0,1)$
is indicated by the large dot.
The entire right boundaries of the plots, $J_y=\infty$, $0<\lambda<\infty$,
are also one-dimensional Heisenberg critical points.
All other points are supposed to have only short-range correlations.
}
\label {fig:simplest_phase_diagrams}
\end{figure}

\begin{figure}
\def\sqr#1#2{{\vcenter{\vbox{\hrule height.#2pt
           \hbox{\vrule width.#2pt height#1pt \kern#1pt
              \vrule width.#2pt}
           \hrule height.#2pt}}}}
\def\square{\mathchoice\sqr54\sqr54\sqr{2.1}3\sqr{1.5}3}
\caption{Estimated values of the critical $\lambda$ as a function
of $J_y$ for the coupled-chain model (\protect\ref{eq:dimerexpHam}).  
From the $\chi$ series, differential approximants displayed are
[2,4;-1] ($\bigcirc$), [3,2;0] ($\square$), and [3,3;-1] ($\diamondsuit$);
from the $M_{rr}$ series, [2,4;-1] ($\triangle$), [3,3;-1] 
($\bigtriangledown$),
and [4,2;-1] ($\triangleright$).
See the text for a discussion
of the significance of the various curves.}
\label {fig:D_planestuff}
\end{figure}

\begin{figure}
\caption{
The symbols correspond to estimates of
the critical $\lambda$ for {\it two\/} chains, corresponding to the
same approximants as the preceding figure (but for $M_{xx}$ rather
than $M_{rr}$).  The curves are discussed in the text.}
\label {fig:twochains}
\end{figure}


\begin{references}

\bibitem{Anderson}
P. W. Anderson, Phys. Rev. Lett. {\bf 64}, 1839 (1990).

\bibitem{Sakai1}
T. Sakai and M. Takahashi,
J. Phys. Soc. Japan {\bf 58}, 3131 (1989).

\bibitem{Azzouz1}
M. Azzouz,
Phys. Rev. B {\bf 48}, 6136 (1993).


\bibitem{Parola1}
A. Parola, S. Sorella and Q. F. Zhong,
Phys. Rev. Lett. {\bf 71}, 4393 (1993).

\bibitem{Scalapino1}
D. J. Scalapino, Y. Imry, and P. Pincus,
Phys. Rev. B {\bf 11}, 2042 (1975).

\bibitem{Chui}
S. T. Chui and M. R. Giri,
Phys. Lett. A {\bf 128}, 49 (1988);
a summary of all numerical work to date is provided
in Fig.~1 of B. Chattopadhyay and S. R. Shenoy,
Phys. Rev. Lett. {\bf 72}, 400 (1994).

\bibitem{Hikami}
S. Hikami and T. Tsuneto, Prog. Theor. Phys. {\bf 63}, 387 (1980).

\bibitem{Kosterlitz}
J. M. Kosterlitz and D. J. Thouless, in
{\sl Progress in Low Temperature Physics} Vol.~7B,
edited by D. F. Brewer (North-Holland, Amsterdam, 1978).

\bibitem{Affleck1}
For a review, see I. Affleck, in {\sl Fields, Strings, and Critical Phenomena},
edited by E. Brezin and J. Zinn-Justin (Elsevier, Amsterdam, 1990).

\bibitem{Haldane}
F. D. M. Haldane, Phys. Rev. Lett. {\bf 61}, 1029 (1988).

\bibitem{twochain_num}
T. Barnes, E. Dagotto, J. Riera, and E. S. Swanson,
Phys. Rev. B {\bf 47}, 3196 (1993).

\bibitem{White1}
S. R. White, R. M. Noack, and D. J. Scalapino,
preprint cond-mat/9403042.

\bibitem{frustratedAF}
For analytical arguments see, for example, N. Read and S. Sachdev,
Phys. Rev. Lett. {\bf 62}, 1694 (1989); {\it ibid.} {\bf 66}, 1773 (1991).
For numerical studies which address the issue of nonmagnetic order
in frustrated $S=1/2$ antiferromagnets see
H. J. Schulz and T. A. L. Ziman, Europhys. Lett. {\bf 18}, 355 (1992);
E. Dagotto and A. Moreo, Phys. Rev. B {\bf 39}, 4744 (1989),
and Phys. Rev. Lett. {\bf 63}, 2184 (1989);
R. R. P. Singh and R. Narayanan, Phys. Rev. Lett. {\bf 65}, 1072 (1990);
M. P. Gelfand, Phys. Rev. B {\bf 42}, 8206 (1990);
M. P. Gelfand, R. R. P. Singh and D. A. Huse, Phys. Rev. B {\bf 40} 10801 (1989).

\bibitem{GSH_expansionreview}
M. P. Gelfand, R. R. P. Singh, and D. A. Huse,
J. Stat. Phys. {\bf 59}, 1093 (1990).

\bibitem{IDA}
M. E. Fisher and H. Au-Yang, J. Phys. A {\bf 12}, 1677 (1979);
D. L. Hunter and G. A. Baker, Jr., Phys. Rev. B {\bf 19}, 3808 (1979).

\bibitem{GSLDQAF}
R. R. P. Singh, M. P. Gelfand, and D. A. Huse,
Phys. Rev. Lett. {\bf 61}, 2484 (1988).

\bibitem{twochains}
H. J. Schulz, Phys. Rev. B {\bf 34}, 6372 (1986);
S. P. Strong and A. J. Millis, Phys. Rev. Lett. {\bf 69}, 2419 (1992).

\end{references}
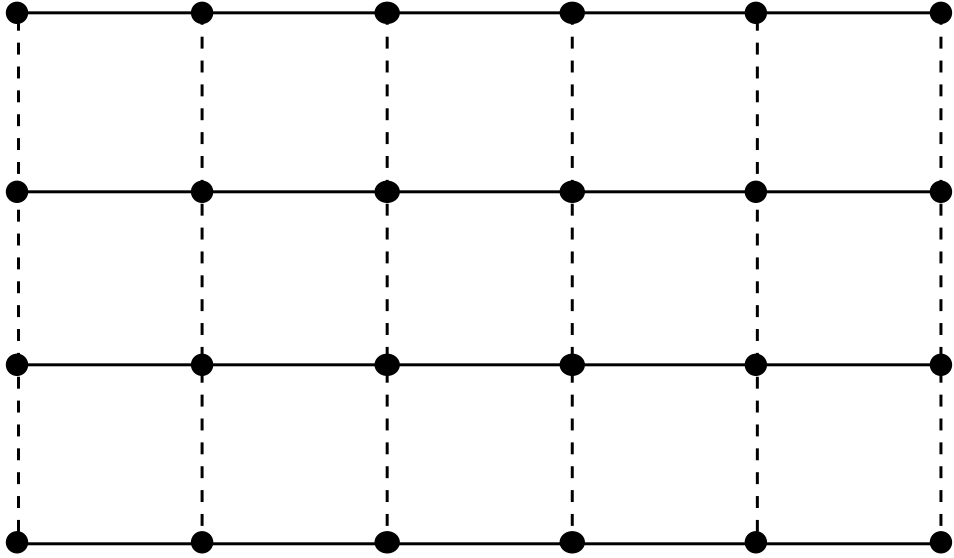
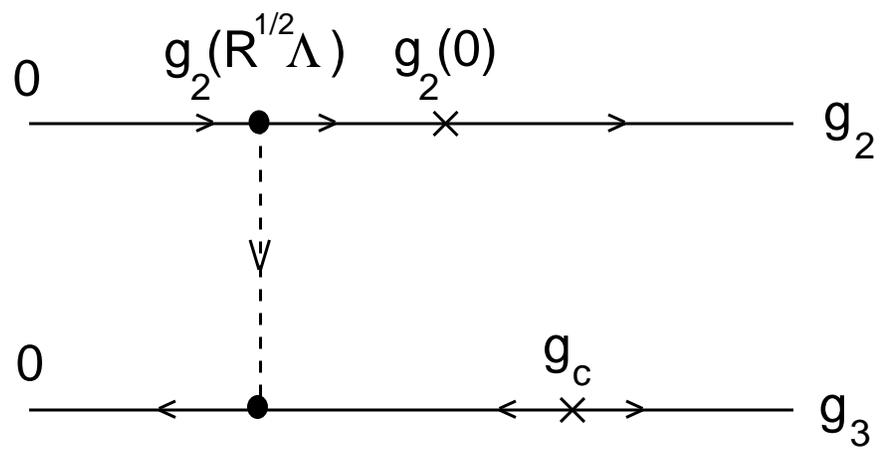
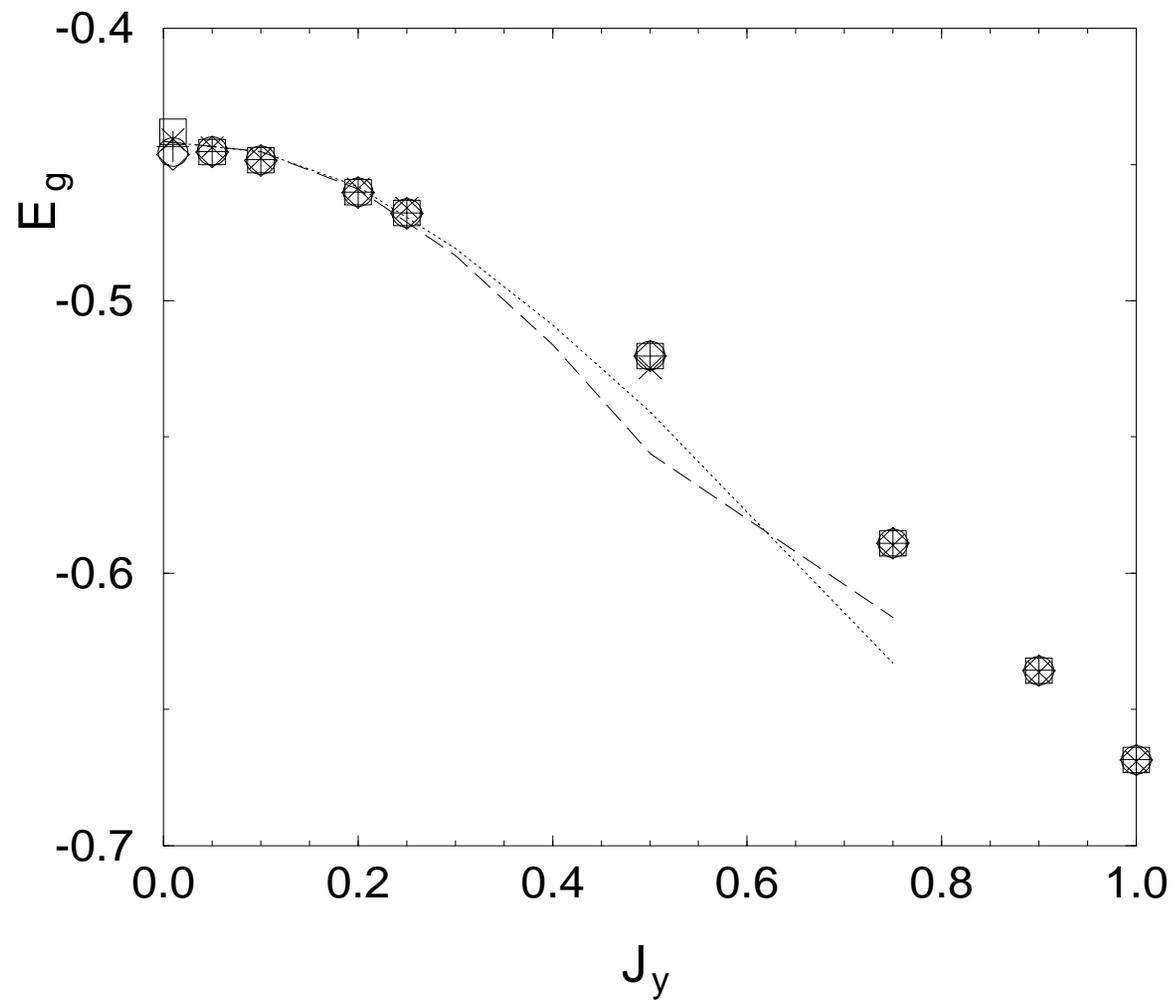
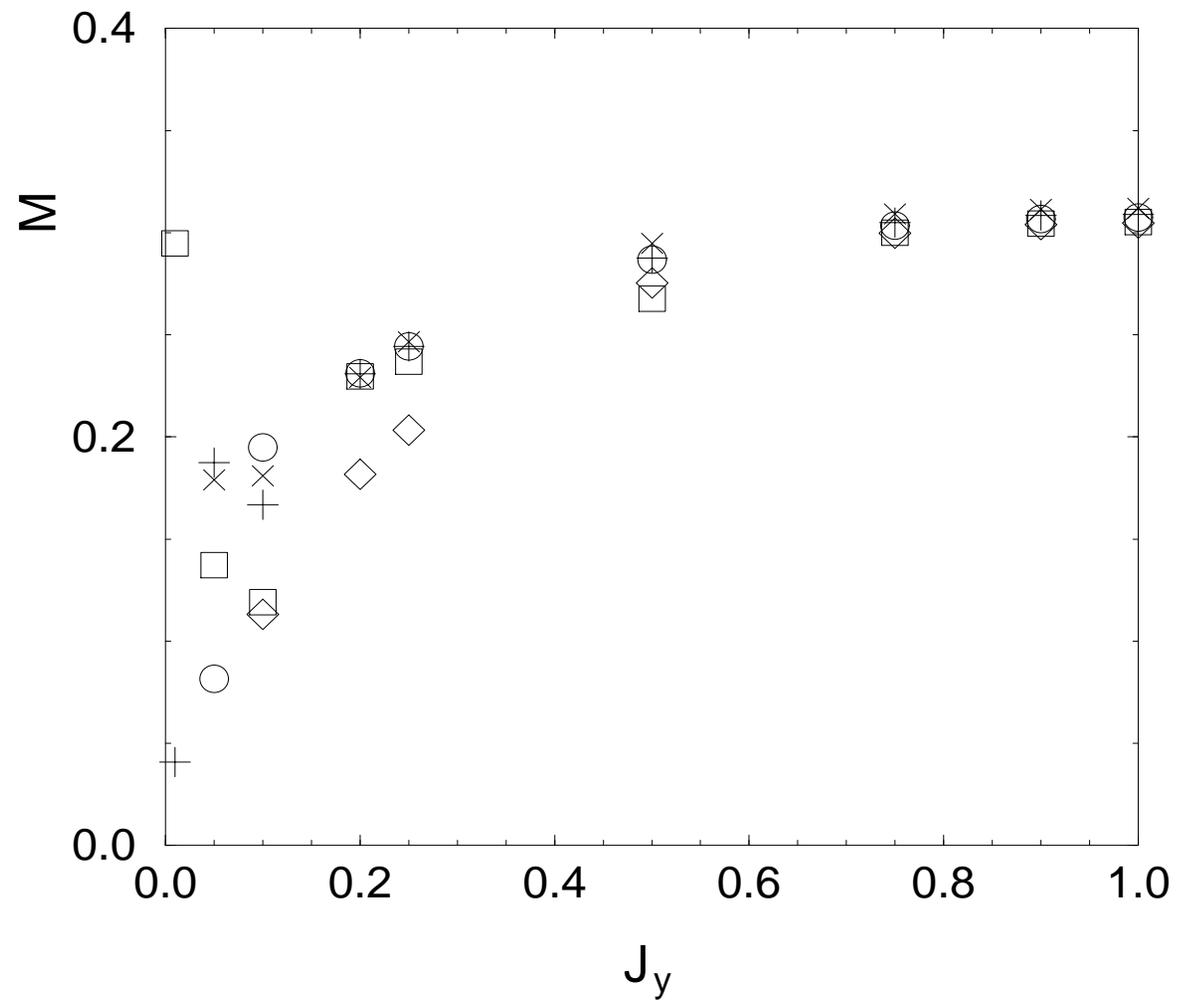
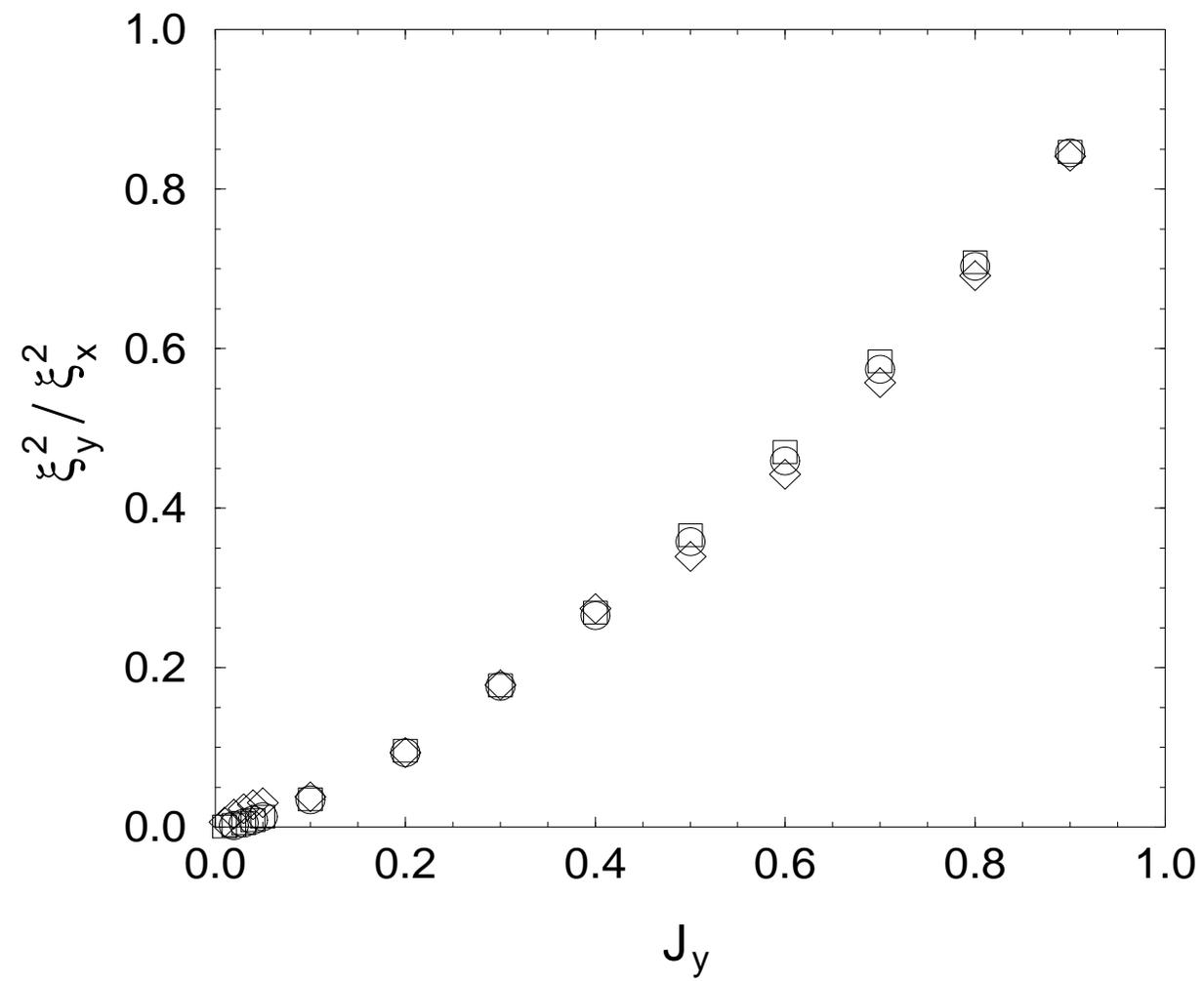
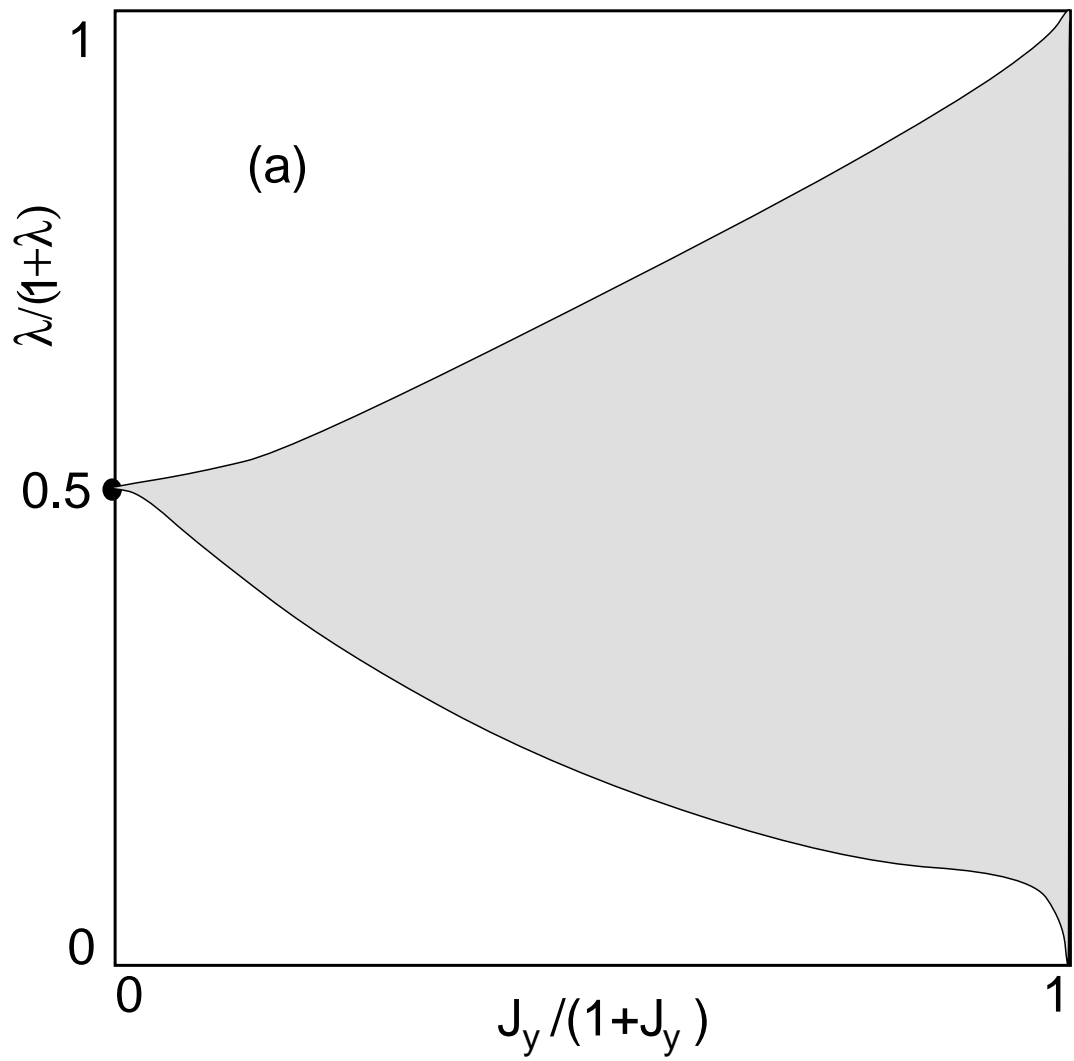
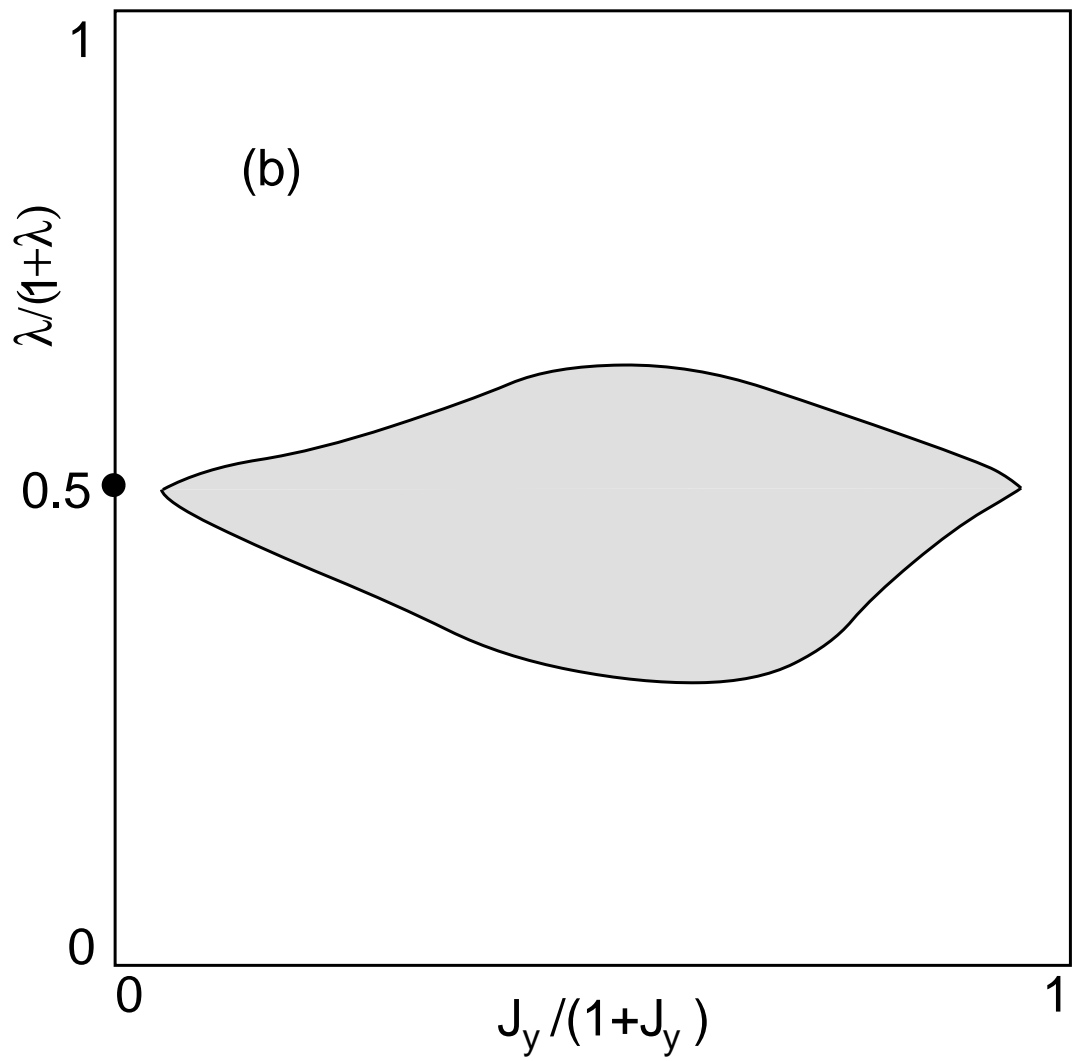
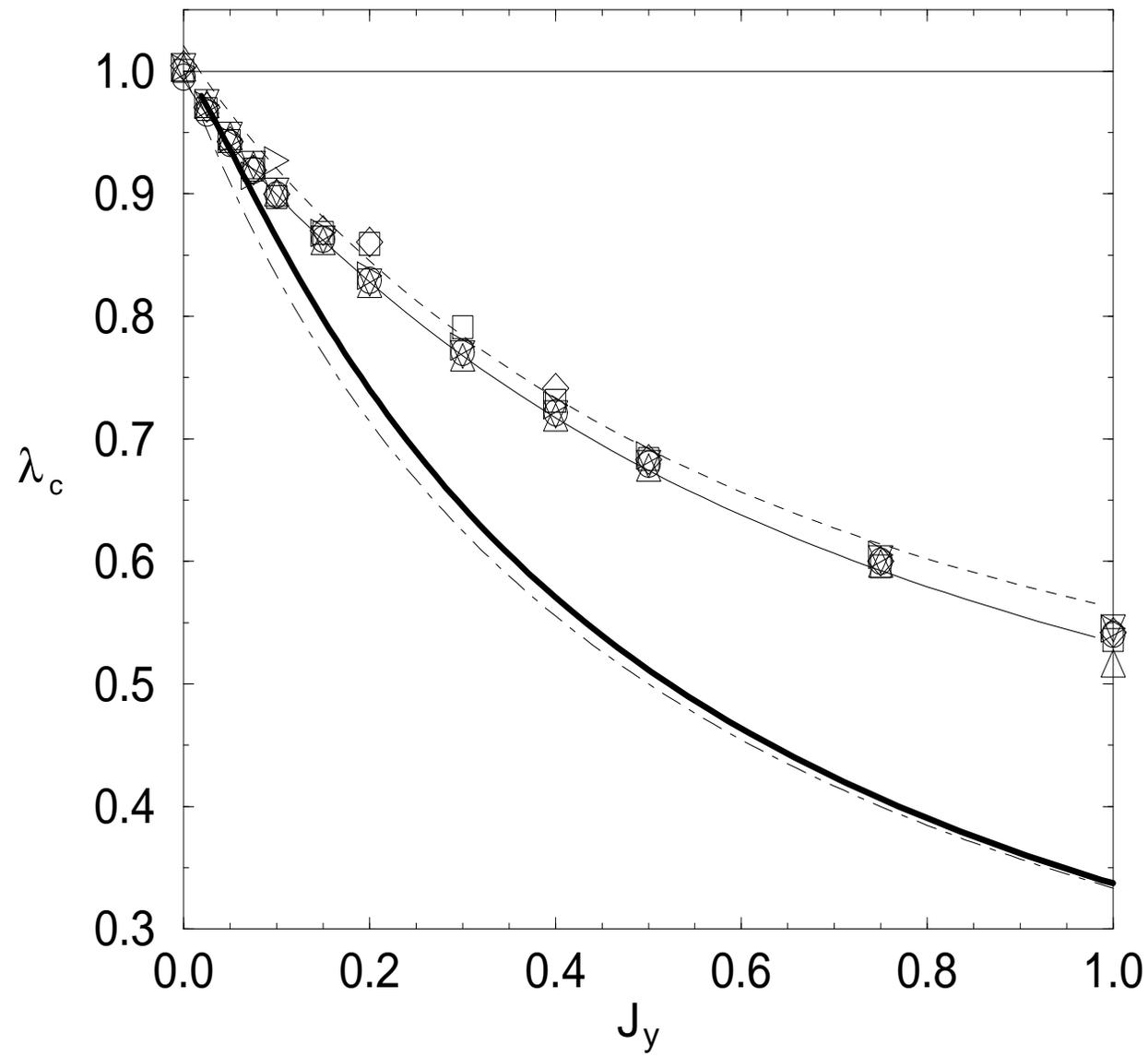
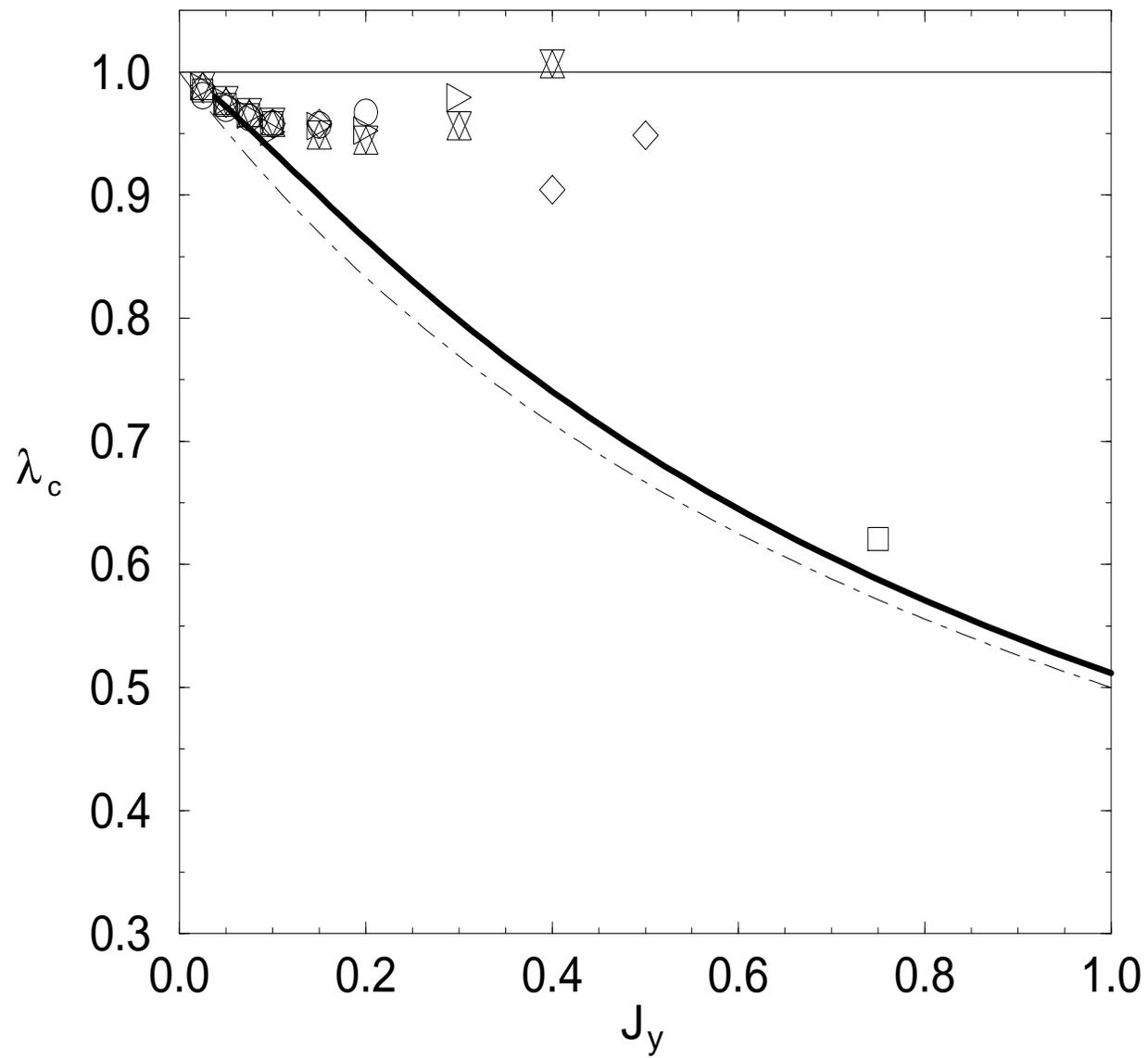
\end{document}